\documentclass[prper,aps,twocolumn,groupedaddress,floats,showpacs,final,superscriptaddress]{revtex4-1}
\usepackage{leading}
\usepackage[latin3]{inputenc}
\usepackage[makeroom]{cancel}
\usepackage{graphicx}
\usepackage{amsmath}
\usepackage{amsfonts}
\usepackage{amssymb}

\usepackage{color}
\usepackage[left=2cm,right=2cm,top=2cm,bottom=2cm]{geometry}
\usepackage{graphicx} 
\usepackage{dcolumn}
\usepackage{bm}
\usepackage{simplewick}
\usepackage{array}
\usepackage{appendix}

\RequirePackage[
   hyperindex,colorlinks,bookmarksnumbered,
   plainpages=true,pdfstartview=FitH]{hyperref}
\hypersetup{linkcolor=blue,urlcolor=blue,citecolor=blue}
\usepackage{hyperref}
\usepackage{mathrsfs}
\definecolor{purple}{rgb}{0.5,0,0.6}

\usepackage{ulem}
\renewcommand{\emph}[1]{\textit{#1}}
\definecolor{darkblue}{rgb}{0,0,0.5}
\definecolor{darkgreen}{rgb}{0,0.5,0}
\definecolor{darkred}{rgb}{.7,0,0}
\definecolor{purple}{rgb}{0.5,0,0.6}
\definecolor{orange}{rgb}{1,0.5,0}
\definecolor{grey}{rgb}{.6,.6,.6}
\definecolor{lightpink}{rgb}{1,0.7,0.75}
\definecolor{pink}{rgb}{1,0.4,0.58}
\definecolor{deeppink}{rgb}{1,0.08,0.58}

\newcommand{\DK}[1]{{\color{black}{#1}}} 

\renewcommand{\emph}[1]{\textit{#1}}

\newcommand{\Tk}{T_{\rm K}}

\newcommand{\ms}{\mathcal{S}}
\newcommand{\mk}{\mathcal{K}}

\begin{document}
\title{
Overscreened Kondo problem with large spin and large number of orbital channels: \\
two distinct semiclassical limits in quantum transport observables}

\author{D. B. Karki}
\affiliation{Division of Quantum State of Matter, Beijing Academy of Quantum Information Sciences, Beijing 100193, China}
\author{Mikhail N. Kiselev}
\affiliation{The  Abdus  Salam  International  Centre  for  Theoretical  Physics  (ICTP),
Strada  Costiera 11, I-34151  Trieste,  Italy}

\begin{abstract}
We investigate the quantum transport through Kondo impurity assuming both a large number of orbital channels $\mathcal K$$\gg $$1$ for the itinerant electrons and a semiclassical spin ${\cal S}$ $\gg $ $1$ for the impurity. The non-Fermi liquid regime of the Kondo problem is achieved in the overscreened sector 
$\mathcal K>2\mathcal S$.  We show that there exist two distinct semiclassical regimes for the quantum transport through impurity: i) 
$\mathcal K$ $\gg$ $\mathcal S$ $\gg$ $1$, differential conductance vanishes and ii) 
$\mathcal S$$/$$\mathcal K$$=$$\mathcal C$ with $ 0$$<$$\mathcal C$$<$$1/2$, differential conductance reaches some non-vanishing fraction of its unitary value. Using conformal field theory approach we analyze behavior of the quantum transport observables and residual entropy in both semiclassical regimes. We show that the semiclassical limit ii) preserves the key features of resonance scattering and the most essential fingerprints of the non-Fermi liquid behavior. We discuss possible realization of two semiclassical regimes in semiconductor quantum transport experiments.
\end{abstract}

\date{\today}
\maketitle

\paragraph*{Introduction.}
The paradigmatic {\color{black} phenomenon} of Kondo screening~\cite{Kondo} successfully provides the understanding of a plethora of problems associated with the physics of strongly correlated electron systems~\cite{Nozieres}. Classification of the Kondo effects originated from the exchange interaction between the localized impurity spin $\ms$ and the itinerant electrons via several conduction channels $\mk$ has been put forwarded by Nozieres and Blandin (NB)~\cite{Nozieres_Blandin_JPhys_1980}. The seminal work of Nozieres~\cite{Nozieres} based on the local Fermi-liquid (FL) theory provides a consistent description for the case of $2\ms\geq\mk$. Different theoretical and numerical techniques have been established~\cite{Wilson, wigmann_JETP(38)_1983, Andrei_RevModPhys_1983,AFFLECK_NPB_1990,Affleck_Lud_PRB(48)_1993} for the study of Kondo effects in non-FL regime associated with the condition $\mk{>}2\ms$.

The celebrated work of Affleck and Ludwig (AL)~\cite{affadd,Affleck_Lud_PRB(48)_1993} based on the boundary conformal field theory (BCFT) is one of the major breakthrough in understanding the multi-channel Kondo (MCK) effects in non-FL (NFL) regime. The asymptotic solution of related problems in MCK {\color{black} has} been also obtained by different 
{\color{black} semiclassical} techniques such as large $\mk$ expansion~\cite{Affleck_Lud_PRB(48)_1993, coleman_andrei, sengupta}. However, to the best of our knowledge, the existing works performed the large $\mk$ expansion keeping impurity spin $\ms$ fixed, that is to say $\mk\to\infty$, $\ms/\mk\to 0$. Yet another interesting {\color{black} semiclassical} limit, which received no attention so far, results from the consideration of $\ms\to\infty$, $\mk\to\infty$ with $\ms/\mk\to{\mathcal{C}}$, $1/2>\mathcal{C}>0$. While the first limit $\ms/\mk\to 0$ was previously explored in detail, the investigation of physics of MCK effects associated with the second {\color{black} semiclassical} limit $\ms/\mk\to{\mathcal{C}}$ is the main focus of present work. 
\paragraph*{Mathematical formulation.}
The MCK effect originally introduced by NB describes the exchange coupling between the $\mathcal{K}$ degenerate channel of spin-1/2 conduction electrons and the impurity with an effective spin $\mathcal{S}$. The corresponding Hamiltonian reads
\begin{equation}\label{aama1}
\mathcal{H}{=}{\sum_{k}}\varepsilon_k \left(\psi_k^{\alpha, i}\right)^{\dagger}\psi_{k\alpha, i}{+}J{\bf \mathcal{S}}\cdot\sum_{kk'}\left(\psi_k^{\alpha, i}\right)^{\dagger}
\frac{\sigma_{\alpha}^{\beta}}{2}\psi_{k'\beta,i},
\end{equation}
where $\alpha, \beta=\uparrow,\downarrow$ are the spin and $i{=}1, 2,\cdots\mathcal{K}$ stand for spin and channel indices respectively and $\sigma_{\alpha}^{\beta}$ are the Pauli matrices acting in spin sector. The operator $\psi_{k\alpha,i}$ annihilates an electron in the $k\alpha$ state of the conduction channel $i$. The strength of exchange interaction is accounted for by the parameter $J$. 

Under the s-wave approximation followed by the linearizion of the dispersion relation, the Eq.~\eqref{aama1} is effectively reduced to the one dimensional problem. Describing the left/right moving fermions by the operators $\psi_{\rm L/R, \alpha j}$, the one-dimensional version of the MCK effect in weak coupling regime is given by~\cite{affadd}
\begin{align}\label{amma2}
\mathscr{H}=\!\frac{iv_{\rm F}}{2\pi}\!\!\int^{\infty}_0 \!\!\!\!\!\!dx\Big[&\psi^{\dagger}_{\rm L, \alpha j}(x)\partial_x \psi^{\dagger}_{\rm L, \alpha j}(x){-}\psi^{\dagger}_{\rm R, \alpha j}(x)\partial_x \psi^{\dagger}_{\rm R, \alpha j}(x)\Big]\nonumber\\
&+v_{\rm F}\lambda \psi^{\rm \alpha i}_{\rm L}(0)^{\dagger}\frac{\sigma_{\alpha}^{\beta}}{2}\psi_{\rm L, \beta j}(0)\cdot\mathcal{S},
\end{align}
with the boundary condition {\color{black} (BC)} $\psi_{\rm R, \alpha j}(0)=\psi_{\rm L, \alpha j}(0)$. In Eq.~\eqref{amma2}, $v_{\rm F}$ stands for the Fermi velocity, $\lambda\equiv \nu J$ for $\nu$ being the density of states at the Fermi {\color{black} level}. For the following discussion we consider $v_{\rm F}=1$ and always restrict ourselves in the overscreened situation $\mathcal{K}>2\mathcal{S}$. 

\paragraph*{Strong coupling description.}
The strong coupling description of FL case can be trivially accounted for by simple change in the BC such that $\psi_{\rm R, \alpha j}(0)=-\psi_{\rm L, \alpha j}(0)$ resulting in the $\pi/2$ phase shit between incoming and outgoing electron states. The description of corresponding BC in the NFL regime can be done straightforwardly in terms of the method of BCFT developed by AL. One of the remarkable result of AL BCFT is the exact calculation of the equal-time correlation function of left and right moving fermions providing the general expression of single particle scattering amplitude $\mathbb{S}$~\cite{affadd, Affleck_Lud_PRB(48)_1993} 
\begin{equation}\label{aama3}
\mathbb{S}=\cos \left[\frac{ \pi  (2 \mathcal{S}+1)}{\mathcal{K}+2}\right]\Big/
\cos \left[\frac{\pi}{\mathcal{K}+2}\right].
\end{equation}

\begin{figure}[t]
\includegraphics[width=50mm]{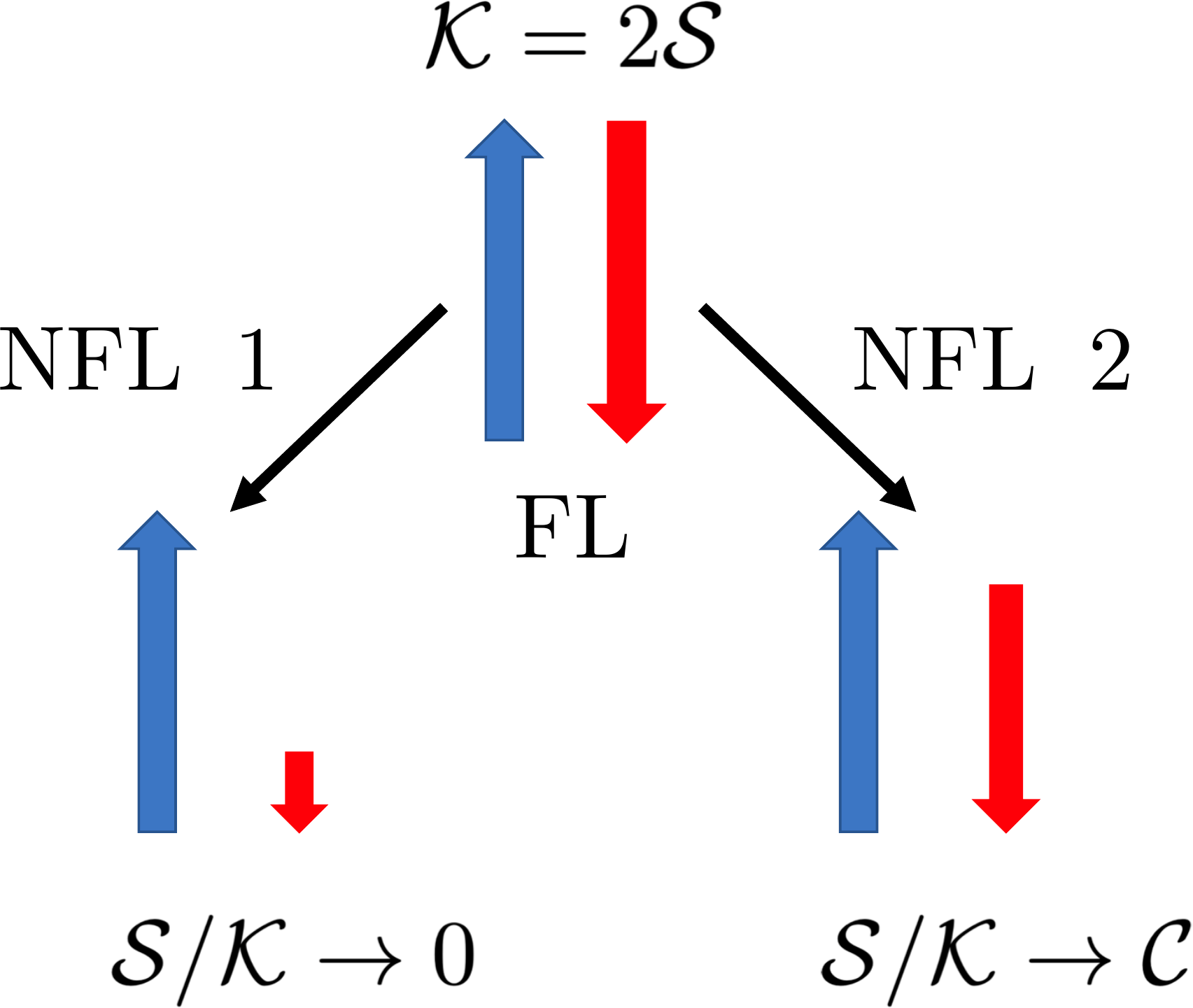}
\vspace*{-3mm}
\caption{Schematic representation of two different semiclassical limits of the overscreened Kondo problem describe by $\mk$ conduction channels and arbitrary spin amplitude $\ms$ (large $\mathcal K$ is denoted by black color, large $\mathcal S$ is illustrated by red color). The condition $\mk=2\ms$ defines fully screened Kondo regime characterized by Fermi liquid behaviour. Two distinct semiclassical limits corresponds to largely overscreened model $\ms/\mk\to 0$ (left) versus {\color{black} weakly overscreened model $\ms/\mk\to \mathcal{C}$ for $\mathcal{C}\to 1/2$} (right). See text for details. }\label{capfig0}
\end{figure}
\begin{figure}[t]
\includegraphics[scale=1.3]{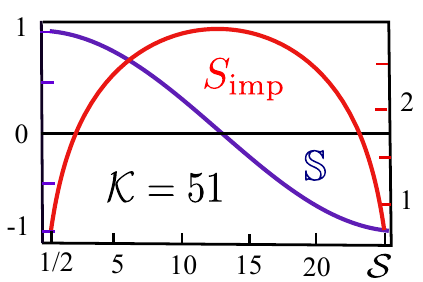}
\vspace*{-3mm}
\caption{The behavior of single particle scattering amplitude $\mathbb{S}$ and the residual impurity entropy $S_{\rm imp}$ with varying the size of the impurity spin $\ms=1/2-25$ for fixed numbers of conduction channel $\mk=51$ in overscreened Kondo effects. Each plots can be further classified into three parts corresponding to the three distinct limits of multi-channel Kondo effects in the overscreening region (see text for the details).}\label{capfig1}
\end{figure}

\vspace*{-3mm}
The zero energy limit of the transport properties in MCK are solely governed by the scattering amplitude $\mathbb{S}$. As seen form the scattering matrix Eq.~\eqref{aama3}, $\mathbb{S}$ vanishes for the special setup satisfying the condition $\mathcal{K}=4\mathcal{S}$ which amounts to the complete absence of single-particle transport such as the case $\mk=2$ and $\mathcal{S}=1/2$ of extreme experimental interests~\cite{Potok_NAT(446)_2007}. Another interesting observation of $\mathbb{S}$ follows that, it becomes positive for $4\mathcal{S}<\mathcal{K}$ and remains negative for the situation $4\mathcal{S}>\mathcal{K}$ both satisfying the condition of the overscreening $\mathcal{K}>2\mathcal{S}$. The later case of $2\mathcal{S}>\mathcal{K}/2>\mathcal{S}$ results in a rather non-trivial property that the scattering amplitude asymptotically becomes $\mathbb{S}\to -1$. These situations have been schematically shown in {\color{black} Fig.~\ref{capfig0} and Fig.~\ref{capfig1}}. Since the amount of coherent transport at zero energy or the conductance is quantified by the quantity $1-\mathbb{S}$, these three cases of overscreened Kondo effect are practically very different. Given that, we propose three further different cases of overscreened Kondo effects: i) $\mathcal{K}=4\mathcal{S}$ with $\mathbb{S}=0$, ii) $4\mathcal{S}<\mathcal{K}$ with $\mathbb{S}>0$ and iii) $2\mathcal{S}>\mathcal{K}/2>\mathcal{S}$ with $\mathbb{S}<0$. 

Another exact result from AL BCFT is the expression of residual impurity entropy $S_{\rm imp}$~\cite{affadd1}
\begin{equation}\label{bau1}
 S_{\rm imp}=\ln \Bigg(\sin \left[\frac{ \pi  (2 \ms+1)}{\mk+2}\right]\Big/
\sin \left[\frac{\pi}{\mk+2}\right]\Bigg),
\end{equation}
{\color{black} being bounded} between $\ln\sqrt{2}$ and $\ln(2\ms+1)$. As the scattering amplitude vanishes at the $\mk=4\ms$, the impurity entropy rather attains its maximum value at the same point (see Fig.~\ref{capfig1}) with the maximum value
\begin{equation}\label{bau2}
S_{\rm imp}^{\rm max}=\ln \left[\csc \left(\frac{\pi }{\mk+2}\right)\right].
\end{equation}

In the following, we characterize three distinct regimes of MCK Kondo effect in NFL regime by two different {\color{black} semiclassical} {\color{black} regimes}: i) $\ms/\mk\to 0$ and ii) $\ms/\mk\to{\mathcal{C}}$, $1/2>\mathcal{C}>0$ {\color{black} (see Fig.~\ref{capfig0})}. We note that the first limit $\ms/\mk\to 0$ was previously explored in detail, therefore, here we mainly focus on the description of the second {\color{black} semiclassical} limit $\ms/\mk\to{\mathcal{C}}$.
\paragraph*{Residual quantum entropy.} 
The Eq.~\eqref{bau1} results in the large $\mk$ limit of impurity entropy for fixed $\ms$ as~\cite{Affleck_Lud_PRB(48)_1993, coleman_andrei}
\begin{equation}\label{bau3}
S_{\rm imp}=\ln\Big[(2 \ms+1)\Bigg[1-\frac{2 \pi ^2 \ms (\ms+1) }{3 \mk^2}\Big]\Bigg]+\mathcal{O}\left(\frac{1}{\mk^3}\right),
\end{equation}
for $\mk\to\infty$ with $\ms/\mk\to 0$, $\ms_{\rm imp}=\ln(2 \ms+1)$. This is expected results since, the $\mk\to\infty$ (keeping $\ms$ fixed) results in the unstable zero coupling fixed point. The degeneracy at the zero-coupling fixed point is simply that of the decoupled spin $2\ms+1$.

To unveil another limit $\ms/\mk\to{\mathcal{C}}$ of impurity entropy, we consider the extreme case of $\mk=2\ms+1$ fulfilling the condition $2\mathcal{S}>\mathcal{K}/2>\mathcal{S}$ and $\ms/\mk\to{\mathcal{C}}$. This limit can also be thought of as that corresponds to the weakly overscreened Kondo effects. In the following, we denote the corresponding quantities for $\mk=2\ms+1$ limit by the primed notations. The large $\mk$ limit of the residual entropy in this case reads
\begin{equation}\label{bau4}
S'_{\rm imp}=\ln2-\frac{\pi^2}{2\mk^2}+\mathcal{O}\left(\frac{1}{\mk^3}\right).
\end{equation}
The seemingly trivial Eqs.~\eqref{bau3} and~\eqref{bau4} rather posses profound physical difference. While the residual entropy increases with impurity spin in $\ms/\mk\to 0$, $\mk\to\infty$ limit, it saturates at the constant value of $\ln 2$ for $\mk=2\ms+1$. In addition, it appears that the large $\mk$ limit of impurity entropy for $\ms=1/2$ formally coincides with that for $\mk=2\ms+1$ case. 
{\color{black} One can see that the ground state degeneracy in semiclassical limit  $\mk\to\infty$ with $\ms/\mk\to 0$ approaches pure spin degeneracy $g_S=2S+1$ while in the semiclassical limit preserving
$\ms/\mk\to 1/2$ the quantum degeneracy approaches the $\mk\to\infty$ $S=1/2$ value $g_{\mk\to\infty, \ms=1/2}=2$.
}

\paragraph*{\DK{Charge transport.}} The zero temperature conductance of MCK effect is expressed in terms of the scattering amplitude~\cite{Affleck_Lud_PRB(48)_1993}
\begin{equation}\label{bau5}
G(T=0)=\frac{2e^2}{h}\;\frac{1-\mathbb{S}}{2}.
\end{equation}
For the qualitative understanding of the conductance in the semi-classical regime of MCK, we perform the large $\mk$ expansion of $\mathbb{S}$ in the above mentioned two different limits ($\ms/\mk\to 0$ and $\mk=2\ms+1$) resulting in
\begin{align}\label{bau6}
\mathbb{S}&=1-\frac{2 \pi ^2 \mathcal{S} (\mathcal{S}+1)}{\mathcal{K}^2}+\mathcal{O}\left(\frac{1}{\mathcal{K}^3}\right),\\
\mathbb{S}'&=-1+\frac{3 \pi ^2}{2 \mk^2}+\mathcal{O}\left(\frac{1}{\mk^3}\right)\label{bau7}.
\end{align}
The Eqs.~\eqref{bau5},~\eqref{bau6} and~\eqref{bau7} then imply that the $\mk$$\to $$\infty$ transport is blocked in the limit of $\ms$$/$$\mk$$\to $$0$ while reaches the unitary value in the other limit of 
$\mk$$=$$2$$\ms$$+$$1$,
\begin{align}
&\left.G(T=0)\right|_{\mk\to\infty}  =0,\nonumber\\
&\left.G'_0  \equiv G'(T=0)\right|_{\mk=2\ms+1\to\infty}=\frac{2e^2}{h}.
\end{align}
{\color{black} The remarkable difference in behavior of differential conductance is related to the sign inversion in the leading term of the scattering matrix expansion (\ref{bau7}), see also Fig. \ref{capfig1}. 
}

The leading $T$-dependence of transport properties in MCK, follows from the perturbation theory in the leading irrelevant coupling constant $\lambda$ at the low energy fixed point ~\cite{Affleck_Lud_PRB(48)_1993}. AL then showed that the NFL BC associated with overscreened MCK effect results in the dimension $\Delta$ of the leading operator (LIO) $\lambda$ at the fixed point
{\color{black}
\begin{equation}\label{aama4}
\Delta=\frac{2}{2+\mathcal{K}}\to \frac{2}{\mathcal K}+ \mathcal{O}\left(\frac{1}{\mk^2}\right).
\end{equation}}
The scaling argument then defined the low energy coupling constant as $\lambda\simeq\pm1/\Tk^{\Delta}$ (either signs are possible depending on the details of the model). The finite temperature ($T\ll\Tk$) conductance of MCK effect is then given by~\cite{Affleck_Lud_PRB(48)_1993}
\begin{equation}\label{amma5}
G(T)=\frac{2e^2}{h}\frac{1}{2}\Big[1-\mathbb{S}-\lambda\;\mathscr{P}(\Delta, \ms) (2\pi T)^{\Delta}\Big],
\end{equation}
where $\mathscr{P}(\Delta, \ms)=2\mathscr{N}\sin(\pi\Delta) \mathscr{C}(\Delta)$. The $\Delta$ dependent parameter $\mathscr{C}(\Delta)$ has recently been computed in Ref.~\cite{DKMK1} and AL provided the parameter $\mathscr{N}$ which depends on $\Delta$ as well as the impurity spin $\mathcal{S}$
\begin{align}\label{aama6}
\mathscr{N}^2\!&=\!\frac{9}{8}\frac{ \Gamma \left(\frac{\mathcal{K}}{\mathcal{K}+2}\right)^2 \left[\cos \left(\frac{2 \pi }{\mathcal{K}+2}\right)-\cos \left(\frac{2 \pi  (2 \mathcal{S}+1)}{\mathcal{K}+2}\right)\right]}{ \left[\!2\!\cos \left(\frac{2 \pi }{\mathcal{K}{+}2}\right){+}1\right]\!\! \left[\!\cos \left(\frac{\pi }{\mathcal{K}{+}2}\right)\!\!\Gamma\! \left(\frac{\mathcal{K}{+}1}{\mathcal{K}{+}2}\right) \!\!\Gamma\! \left(\frac{\mathcal{K}-1}{K+2}\right)\!\right]},\nonumber\\
&\;\;\;\;\;\;\;\;\;\;\;\;\;\;\mathscr{C}(\Delta)=-\frac{1}{\Delta(1+\Delta)}.
\end{align}

Similar to the previous discussion of the residual entropy, we now perform the large $\mk$ expansion of finite temperature {\color{black} differential} conductance for two different limits: $\ms/\mk\to 0$ and $\mk=2\ms+1$ to obtain
\begin{align}\label{amma8}
&G(T)\left|_{\ms/\mk\to 0}\right.{=}\frac{e^2}{h}\!\Big[2\pi^2\!\sqrt{3\ms (\ms{+}1)}\!\left(\frac{2\pi T}{\Tk}\right)^{\!\Delta}\!\!\left(\frac{1}{\mk}{-}\frac{4}{\mk^2}\right)\Big],\nonumber\\
&G'(T)\left|_{\ms/\mk\to \frac{1}{2}}\right.{=}G'_0{+}\frac{e^2}{h}\Big[3\pi^2\!\left(\!\frac{2\pi T}{\Tk}\right)^{\!\Delta}\!\!\left(\!\frac{1}{\mk}{-}\frac{4}{\mk^2}\!\right)\!\Big].
\end{align}
The Eq.~\eqref{amma8} shows {\color{black} a non-trivial} equality relation {\color{black} $\delta G(T, \ms=1/2)=G'(T)-G'_0$} implying that the finite temperature corrections in these two limits are qualitatively different. As an additional note, we stress that the behavior of thermal transports (such as thermopower, thermo-electric conductance) as described in Ref.~\cite{DKMK1} are also different in these two limit.

\paragraph*{\DK{Thermoelectric transport.}} \DK{Thermoelectric transport measurement at nanoscale often provides invaluable informations that cannot be achieved by
charge transport measurements~\cite{casti, *dee1, *last2}. In addition, it is commonly believed that thermopower (Seebeck coefficient) is directly connected to the entropic heat production via the Kelvin formula~\cite{lla}. While the later is well accepted fact for general situation of FL ground state, it is still an ongoing research to find if there exists some relation between entropy and thermopower in the case of NFL ground state~\cite{GM, *GM1}. Calculation of thermopower in MCK effect will thus sheds some light on this topic.

So far our discussion was based on the explicit particle-hole (PH) symmetric low energy model of transport in MCK effects. Such symmetry, however, results in vanishing thermopower. The explicit consideration of potential scattering~\cite{DKMK1} in MCK problems, gives rise to the asymmetric spectral function which provides the finite energy transport in the system. By considering the constant phase shift $\delta_{\rm P}$ produced by the potential scattering, the leading temperature dependence of thermopower $\mathcal{S}_{\rm th}$ has been calculated in Ref.~\cite{DKMK1}:
\begin{equation}\label{yamk1}
\mathcal{S}_{\rm th}=\lambda\mathscr{M}(\Delta, \mathcal{S})\sin2\delta_{\rm P}(2\pi T)^{\Delta},
\end{equation}
where $\mathscr{M}(\Delta, \ms)=2\mathscr{N}\pi\sin(\pi\Delta) \mathscr{D}(\Delta)$ with $\mathscr{D}(\Delta)=1/(1-\Delta)(2+\Delta)$. From above equation, it is clearly seen that the in the semi-classical limit $\ms/\mk\to 0$, $\mathcal{S}_{\rm th}$ still depends on $\mathcal{S}$ (as similar to the conductance behavior). The non-trivial limit of $\mk=2\ms+1$, however, results in the thermopower which is qualitatively the same as that of $\ms=1/2$ with $\mk\to\infty$.

In addition, from our previous discussion, we see that MCK effects always posses finite residual entropy (zero temperature contribution). The thermopower as expressed in Eq.~\eqref{yamk1} certainly vanishes for $T\to 0$. This observation clearly indicates that the thermopower and entropy of a NFL state {\color{black} are not interrelated} in such as direct way as in FL case~\cite{lla}. It has been explicitly shown that all the transport quantities (including thermo-electric {\color{black} power}) of MCK impurity explicitly depend on the magnitude of the impurity spin $\ms$ and the numbers of conduction channels $\mk$~\cite{DKMK1}. The two different {\color{black} semiclassical} limits can then be investigated in the analogous way as done for charge conductance above.
\paragraph*{\DK{Thermodynamics.}}
From the AL BCFT, it can be straightforwardly seen that the thermodynamic measure of impurity specific heat $C_{\rm imp}$ can be solely described in terms of $\mk$ which is independent of $\ms$~\cite{AFFLECK_AP_1995}. The investigation of two different {\color{black} semiclassical} limits in the spirit of seminal works~\cite{WT1,*WT2,*WT3} for other thermodynamical parameters of MCK impurities (including behavior in external magnetic field)} can thus be an interesting future project.

\paragraph*{Discussion.}
{\color{black} Two distinct semiclassical limits can be experimentally accessed through quantum transport measurements in semiconductor nano-devices. The limit
$\mathcal K$$\gg $$1$ can be engineered in the setup consisting of a small {\color{black} lateral} quantum dot
(single-electron-transistor) surrounded by $\mathcal K $$-$$ 1$ large metallic quantum dots 
(metallic droplets) controlled by independent gates  (see detailed discussion in \cite{DKMK1}). The condition $\mathcal S$ $\gg$ $1$
can be achieved by various methods. First suggestion is based on replacement of the small semiconductor quantum dot by a set of small capacitively coupled quantum dots such a way that the large spin multiplet $\mathcal S$ $\gg$ $1$ is separated from the smallest spin multiplet (either singlet or doublet) by an energy gap being smaller compared to the Kondo temperature. Second suggestion is to use the Fock-Darwin states~\cite{kka} of a small semiconductor quantum dot for achieving the large spin configuration. {\color{black} The degeneracy of the multi-orbital states of the Fock-Darwin atom can be adjusted by external magnetic field 
\cite{addsg1, *addsg2}. Spins of electrons occupying different orbitals are aligned in accordance with the Hund's rule facilitating the large $\mathcal S$ configuration}. {\color{black} While NFL regimes in quantum transport have not yet been experimentally accessible with vertical QDs, achieving it remains a challenging problem. Usage of surrounding vertical dots or metallic droplets similarly to lateral dots experiments can be an interesting direction to try.} The semiclassical limit 
$\mathcal S$$/$$\mathcal K$$=$$\mathcal C$ with $\mathcal C $$\to $$1$$/$$2$} can be controlled {\color{black} in Fock-Darwin atom} by proper balance between number of metallic droplets and the quantum dot spin. The manifestation of pronounced non-Fermi-liquid behavior in transport observables preserved in the semi-classical regime is the key prediction of this work.


\paragraph*{\DK{Conclusion.}} Based on the Affleck-Ludwig boundary conformal field theory for multi-channle $\mk$ overscreened Kondo effects with arbitrary impurity spin $\ms$, we showed that there exists two different {\color{black} semiclassical limits} resulting in two strikingly different conclusion. Namely we demonstrated that the commonly studied trivial {\color{black} semiclassical} limit $\mk{\to}\infty$, $\ms/\mk{\to} 0$ and the non-trivial counterpart $\ms{\to}\infty$, $\mk{\to}\infty$ with $\ms/\mk{\to}{\mathcal{C}}$, $1/2{>}\mathcal{C}{>}0$ are two very different limits. While the former case with $\mk{\to}\infty$ corresponds to the vanishing transport, the corresponding later case amounts to the complete coherent transport there by reaching the unitary conductance. Considering these observations, we proposed three different classes of the overscreened Kondo effects $\mk{>}2\ms$: i) $\mathcal{K}{=}4\mathcal{S}$ with vanishing scattering amplitude $\mathbb{S}{=}0$, ii) $4\mathcal{S}{<}\mathcal{K}$ with $\mathbb{S}{>}0$ and iii) $2\mathcal{S}{>}\mathcal{K}/2{>}\mathcal{S}$ with $\mathbb{S}{<}0$. The interesting interplay of residual entropy in these three classes of overscreened Kondo effects has been investigated. We believe that the proposed two ways of doing {\color{black} semiclassical} calculation in multi-channel Kondo effects would be of paramount importance for bench-marking the numerical results and theoretical calculations. Although, in this work we just showed the phenomenology of two different {\color{black} semiclassical} approaches with an explicit consideration of thermo-electric transport and residual entropy, the corresponding investigations for other quantities such as heat transport, finite temperature coherent transport and thermodynamics remain valid avenues for future research. 
\paragraph*{Acknowledgments.} We are indebted to Paul Wiegmann for drawing our attention to the existence of two distinct semiclassical limits in the thermodynamic properties of the overscreened Kondo problem accessible by the Bethe-ansatz and Natan Andrei for illuminating discussions. The work of M.K.  is conducted within the framework of the Trieste Institute for Theoretical Quantum Technologies (TQT).
\vspace*{-3mm}
\end{document}